\documentstyle[preprint,aps,12pt]{revtex}
\tightenlines
\begin{document}
\draft
\preprint{YITP-98-26,\\hep-ph/9804342}

\title{
$K \rightarrow \pi\pi$, $K_L$-$K_S$ mass difference \\and the
Dalitz decays of $K_L$\footnote{Talk given at the Workshop on Fermion 
Mass and CP Violation, March 5 -- 6, 1998, Hiroshima International
Cooperation Center, Higasi-Hiroshima}
}
\author{K. Terasaki\\ Yukawa Institute for Theoretical Physics,\\
Kyoto University, Kyoto 606-8502, Japan
}
\date{March 5th , 1998 }
\maketitle
\thispagestyle{empty}
\begin{abstract}
$K \rightarrow \pi\pi$, $K_L$-$K_S$ mass difference, 
$K_L \rightarrow \gamma\gamma$ and the Dalitz decays of $K_L$ are
studied systematically by assuming that their amplitude can be
described in terms of a sum of short distance and long distance
contributions. Dominance of the short distance effect on the 
$K_L$-$K_S$ mass difference will be checked by the Dalitz decays of 
$K_L$ in a way consistent with the $K \rightarrow \pi\pi$ and 
$K_L \rightarrow \gamma\gamma$ decays. 
\end{abstract}

\section{Introduction}

Our starting point to study nonleptonic weak processes is to assume
that their amplitude can be decomposed into a sum of 
{\it short distance} and {\it long distance} terms and that 
the long distance amplitude is dominated by dynamical contributions 
of various hadrons\cite{Terasaki-B}. 

It has been known that a short distance contribution is small in the 
$K_L \rightarrow \gamma\gamma$ decay\cite{Gaillard-Lee} and a naively 
factorized $\Delta I=1/2$ amplitude for the $K \rightarrow \pi\pi$
decays which is now classified into the short distance one is also
much smaller than the observed one\cite{BBL}. However, importance of 
long distance contribution to the $K_L$-$K_S$ mass difference, 
$\Delta m_K = m_{K_L} - m_{K_S}$, is still in controversy although 
the mass difference has been used to test theories within or beyond 
the standard model by assuming explicitly or implicitly dominance of 
short distance contribution. Therefore, it will be meaningful to 
study a role of the long distance contribution in $\Delta m_K$ and 
test it in some other processes. To this, we will study the following 
two extreme cases since theoretical and experimental ambiguities are 
still large: (i) the short distance contribution vanishes, 
$(\Delta m_K)_{\rm SD}=0$, and (ii) the long distance contribution 
vanishes, $(\Delta m_K)_{\rm LD}=0$. Then we investigate responses 
of the Dalitz decays of $K_L$ in the above two extreme cases. 

Before we study amplitudes for the weak processes mentioned above, 
we review briefly the effective weak Hamiltonian which is usually 
written in the form\cite{HW,SVZ},  
\begin{equation}
H_w = {G_F \over \sqrt{2}}V_{ud}V_{us}\Bigl\{c_1O_1 + c_2O_2 + 
({\rm penguin})\Bigr\} + h.c.,                        \label{eq:HW}
\end{equation}
or equivalently, 
\begin{equation}
H_w = {G_F \over \sqrt{2}}V_{ud}V_{us}\Bigl\{c_-O_- + c_+O_+ + 
({\rm penguin})\Bigr\} + h.c.                     \label{eq:HW-pm}
\end{equation}
with $O_{\pm} = O_1 \pm O_2$, where the four quark operators $O_1$ 
and $O_2$ are given by 
\begin{equation}
O_1 = :(\bar us)_{V-A}(\bar du)_{V-A}: \quad {\rm and} \quad 
O_2 = :(\bar uu)_{V-A}(\bar ds)_{V-A}: .
\end{equation}
$O_\pm$ transform like ${\bf 8_a}$ and {\bf 27} of the flavor
$SU_f(3)$ and are responsible for the $\Delta I = 1/2$ and $3/2$ 
amplitudes for the $K \rightarrow \pi\pi$ decays, respectively. 
$V_{ij}$ denotes a CKM matrix element\cite{CKM} which is taken to be 
real since CP invariance is always assumed in this paper. When we
apply the factorization prescription to the $K \rightarrow \pi\pi$ 
amplitudes later, we use the so-called BSW 
Hamiltonian\cite{BSW,NRSX} 
\begin{equation}
H_w^{BSW} = {G_F \over \sqrt{2}}V_{ud}V_{us}
\Bigl\{a_1O_1^H + a_2O_2^H + ({\rm penguin})^H\Bigr\} + h.c. 
                                                   \label{eq:HW-BSW}
\end{equation}
which can be obtained from Eq.(\ref{eq:HW}) by using the Fierz 
reordering. The operators with the superscript $H$, {\it i.e.}, 
$O_1^H$, $O_2^H$ and $({\rm penguin})^H$, should no longer be Fierz 
reordered. The coefficients $a_1$ and $a_2$ are given by 
\begin{equation}
a_1 = c_1 + {c_2 \over N_c}\simeq 1.14,
\quad a_2 = c_2 + {c_1 \over N_c}\simeq -0.209,
                                                \label{eq:coef-BSW}
\end{equation}
where $N_c$ is the color degree of freedom. Numerical values are
obtained by using the values of $c_1$ and $c_2$ with the leading 
order QCD corrections\cite{QCD-corr}. 

In the next section, we will study two photon decays, 
$K_L \rightarrow \gamma\gamma^{(*)}$, and the Dalitz decays of $K_L$, 
where $\gamma^{(*)}$ denotes an (off-mass-shell) photon. 
The amplitude will be given by two independent matrix elements of 
$H_w$ taken between pseudo scalar meson states and between helicity 
$\lambda = \pm 1$ vector meson states. In {\bf 3}, the $K_L$-$K_S$ 
mass difference will be investigated. Its short distance term 
$(\Delta m_K)_{\rm SD}$ is proportional to 
$\langle{K^0|O_{\Delta S=2}|\bar K^0}\rangle$ arising from the box 
diagrams\cite{Gaillard-Lee} and the long distance one 
$(\Delta m_K)_{\rm LD}$ is dominated by a sum of contributions of 
pseudo scalar and vector meson poles and of $(\pi\pi)$ intermediate 
states\cite{Itzykson}. In {\bf 4}, the $K \rightarrow \pi\pi$ decays 
will be investigated. The short distance amplitude is estimated by 
using the naive factorization prescription\cite{BSW}. The long 
distance amplitude is assumed to be dominated by dynamical 
contributions of various hadron states and is estimated by using a 
hard pion approximation\cite{hard pion,suppl}. Since the naively 
factorized (short distance) amplitude does not satisfy the 
$\Delta I = 1/2$ rule and its $\Delta I = 1/2$ part is much smaller 
than the observed one, the (long distance) hard pion amplitude of 
$\Delta I=1/2$ should be much larger than the factorized one and 
satisfy the $\Delta I = 1/2$ rule to reproduce the observation. 
To realize this, asymptotic matrix elements of $H_w$ taken between 
$\langle{\pi}|$ and $|{K}\rangle$ (or $|{K^*}\rangle$) must satisfy 
the rule, which will be demonstrated by using a simple quark counting 
in {\bf 5}. Asymptotic matrix elements of $H_w$ taken between the 
ground-state-meson states will be parameterized in the same section. 
Inserting the parameterization into the long distance amplitudes, we 
will compare our result with experimental data on 
$K\rightarrow \pi\pi$, $\Delta m_K$, $K_L\rightarrow \gamma\gamma$ 
and the Dalitz decays of $K_L$ in {\bf 6}. In the final section, we 
will provide a brief summary. 

\section{Two photon decays of $K_L$}

Now we study the $K_L \rightarrow \gamma\gamma$ and $\gamma\gamma^*$
decays. We consider Lorentz invariant amplitudes in the infinite 
momentum frame (IMF) for later convenience. As mentioned before, it 
is known\cite{Gaillard-Lee} that short distance contribution to the 
$K_L \rightarrow \gamma\gamma$ is small. Therefore, we neglect it and 
consider only long distance effects which will be dominated by pole 
amplitudes since contributions of two and more pion intermediate
states are suppressed because of the approximate CP invariance and 
small phase space volume, respectively. However a sum of pseudo 
scalar meson ($P=\pi^0,\,\eta,\,\eta'$) pole amplitudes\cite{Pakvasa} 
\begin{equation}
A_P(K_L \rightarrow \gamma\gamma) = 
\sum_{P_i}{\langle{K_L|H_w|P_i}\rangle 
A(P_i \rightarrow \gamma\gamma) \over (m_{P_i}^2 - m_K^2)}
                                                 \label{eq:P-pole}
\end{equation}
with the usual $\eta$-$\eta'$ mixing angle, 
$\theta_P \simeq -20^\circ$\cite{PDG}, is not 
sufficient\cite{D'Ambrosio-Espriu} to reproduce the observed 
rate\cite{PDG}, 
$\Gamma(K_L \rightarrow \gamma\gamma)_{\rm expt} 
= (7.26 \pm 0.35)\times 10^{-12}$ eV. 
Therefore we have to take into account some other contributions. 
Although a possible role of the pseudo scalar glue-ball ($\iota$)
through the penguin effect has been considered in Ref.\cite{Pakvasa}, 
it will be not very important in the present perspective because of 
its high mass and small rate 
$B(\iota \rightarrow \gamma\gamma)_{\rm expt} < 1.2 $ keV. 
Another possible contribution to the $K_L \rightarrow \gamma\gamma$ 
will be the $K^*$ meson pole with the vector meson dominance 
(VMD)\cite{VMD}. However there have been some arguments against 
it\cite{Munczek}. These arguments are based on the field 
algebra\cite{field-algebra} and their weak Hamiltonian consists of 
{\it symmetric} products of left-handed currents and transforms 
like ${\bf 8_s}$ of $SU_f(3)$. It is much different from the standard 
model presented in the previous section. Therefore we should not be 
restricted by such arguments. Since the VMD in the electro-magnetic 
interactions of hadrons can be derived\cite{Bando} independently of 
the field algebra, we now can be free from the above arguments in 
Ref.\cite{Munczek} even if we use the VMD. In this way, we can safely 
take into account the $K^*$ pole contribution in the 
$K_L \rightarrow \gamma\gamma$ decay\cite{Terasaki-FF}. 
Its off-mass-shell amplitude is given by 
\begin{eqnarray}
&& A_{K^*}(K_L \rightarrow \gamma\gamma^*(k^2))
= \sum_{V_i}\sum_{V_j}\sqrt{2}X_{V_i}X_{V_j}
G_{K^0K^{*0}V_i}
\langle{K^{*0}|H_w|V_j}\rangle_{\lambda=\pm1}\nonumber\\
&&\hspace{5cm}\times\Bigl\{ 
{1 \over m_{V_i}^2(m_{K^*}^2-k^2)(m_{V_j}^2-k^2)} 
+ {1 \over (m_{V_i}^2 - k^2)m_{K^*}^2m_{V_j}^2}
\Bigr\} 
                                                 \label{eq:K^*-pole}
\end{eqnarray}
with $V_i = \rho^0$, $\omega$ and $\phi$. 
$X_{V_i}=em_{V_i}^2/f_{V_i}$ is the photon-vector meson coupling 
strength and $f_{V_i}$ is the usual photon-vector meson transition 
moment. The subscript $\lambda = \pm 1$ of the matrix element 
$\langle{K^{*0}|H_w|V_j}\rangle_{\lambda=\pm 1}$ denotes the helicity 
of the vector meson states which sandwich $H_w$. The $K^*$ pole 
amplitude for the $K_L \rightarrow \gamma\gamma$ decay is simply 
obtained by putting $k^2 = 0$ in the above off-mass-shell amplitude, 
Eq.(\ref{eq:K^*-pole}). 

The pseudo scalar meson pole amplitude, Eq.(\ref{eq:P-pole}), can be 
extrapolated into the off-mass-shell region approximately in the 
form, 
\begin{equation}
A_P(K_L \rightarrow \gamma\gamma^*(k^2)) = 
\sum_{P_i}{\langle{K_L|H_w|P_i}\rangle 
A(P_i \rightarrow \gamma\gamma) 
\over (m_{P_i}^2 - m_K^2)(1 - k^2/\Lambda_P^2)},
                                                 \label{eq:P-pole*}
\end{equation}
since the observed form factors for the $\pi^0$, $\eta$ and 
$\eta' \rightarrow \gamma\gamma^*$ decays are approximately described 
in the form\cite{Landsberg}, $\sim (1 - k^2/\Lambda_P^2)^{-1}$ with 
$\Lambda_P \simeq m_\rho$. For more precise arguments, however, we 
may have to use a more improved result from recent 
measurements\cite{CLEO}, 
$\Lambda_\pi = 776 \pm 10 \pm 12 \pm 16 \,\,{\rm MeV}$, 
$\Lambda_\eta = 774 \pm 11 \pm 16 \pm 22  \,\, {\rm MeV}$ 
and $\Lambda_{\eta'} = 859 \pm 9 \pm 18 \pm 20 \,\,{\rm MeV}$. 
In this way, the amplitude and the form factor for the 
$K_L \rightarrow \gamma\gamma^*$ are approximately given by  
\begin{equation}
A(K_L \rightarrow \gamma\gamma^*(k^2)) 
\simeq  A_P(K_L \rightarrow \gamma\gamma^*(k^2)) 
+ A_{K^*}(K_L \rightarrow \gamma\gamma^*(k^2))   
                                                 \label{eq:Dalitz-A}
\end{equation}
and 
\begin{equation}
f(k^2) = {A(K_L \rightarrow \gamma\gamma^*(k^2)) 
\over A(K_L \rightarrow \gamma\gamma)}, 
                                                 \label{eq:Dalitz-FF}
\end{equation}
respectively, where 
\begin{equation}
A(K_L \rightarrow \gamma\gamma) 
= A(K_L \rightarrow \gamma\gamma^*(k^2 = 0)).    \label{eq:2-gamma}
\end{equation}
As seen in Eqs.(\ref{eq:Dalitz-A}) and (\ref{eq:Dalitz-FF}) with 
Eqs.(\ref{eq:P-pole}), (\ref{eq:K^*-pole}) and (\ref{eq:P-pole*}), 
the amplitude $A(K_L \rightarrow \gamma\gamma^*)$ and the form factor 
$f(k^2)$ for the $K_L\rightarrow\gamma\gamma^*$ have been written in 
terms of {\it asymptotic ground-state-meson matrix elements} of $H_w$ 
(matrix elements of $H_w$ taken between the ground-state-meson states 
with infinite momentum). 

Since the Dalitz decay of $K_L$ proceeds dominantly as 
$K_L \rightarrow \gamma\gamma^* \rightarrow \gamma \ell^+\ell^-$, 
its branching fraction is given by the following formula\cite{BMS}, 
\begin{equation}
R_{\gamma\ell^+\ell^-} 
= {\Gamma(K_L \rightarrow \gamma \ell^+\ell^-) 
         \over \Gamma(K_L \rightarrow \gamma \gamma)} 
= \bigl[\Gamma(K_L \rightarrow \gamma \gamma)\bigr]^{-1}
\int_{x_{\rm min}}^{1} dx 
       \Biggl[{d\Gamma(K_L \rightarrow \gamma \ell^+\ell^-)
                                                \over dx}\Biggr], 
                                             \label{eq:Dalitz-rate}
\end{equation}
with $x_{\rm min}= (2m_\ell /m_K)^2$, where 
\begin{eqnarray}
&& \bigl[\Gamma(K_L \rightarrow \gamma \gamma)\bigr]^{-1}
  {d\Gamma(K_L \rightarrow \gamma \ell^+\ell^-) \over dx} 
                                                       \nonumber\\
&& \hspace{2.5cm}
= \Biggl({2\alpha \over 3\pi}\Biggr){(1 - x)^3 \over x}
   \Biggl[1 + 2\Biggl({m_\ell \over m_K}\Biggr)^2{1 \over x}\Biggr]
\Biggl[1 - 4\Biggl({m_\ell \over m_K}\Biggr)^2{1 \over x}
\Biggr]^{1/2}|f(x)|^2.                          \label{eq:Diff-rate}
\end{eqnarray}
In the previous analyses\cite{BMS,Ko} which were restricted by the 
arguments in Ref.\cite{Munczek}, the $K^*$ pole had to vanish in the 
$K_L \rightarrow \gamma\gamma$ while it survived in the Dalitz decays 
of $K_L$. In this case, however, it will be hard to reproduce the 
observed $\Gamma(K_L\rightarrow \gamma\gamma)$ in consistency with the 
$K\rightarrow \pi\pi$ decays if the usual $\eta$-$\eta'$ mixing angle 
$\theta_P\simeq -20^\circ$ is taken, as mentioned before. 

\section{$K_L$-$K_S$ mass difference}

Now we study the $K_L$-$K_S$ mass difference $\Delta m_K$ by 
decomposing it into a sum of {\it short distance} and 
{\it long distance} contributions\cite{Delta-m_K}, 
\begin{equation}
\Delta m_K = (\Delta m_K)_{\rm SD} + (\Delta m_K)_{\rm LD}.  
                                              \label{eq:Delta-m_K}
\end{equation}
The short distance contribution $(\Delta m_K)_{\rm SD}$ is 
proportional to the matrix element of the $\Delta S=2$ box
operator\cite{Gaillard-Lee} taken between $\langle{K^0}|$ and 
$|{\bar K^0}\rangle$, {\it i.e.}, 
\begin{equation}
(\Delta m_K)_{\rm SD} \propto 
\langle{K^0|O_{\Delta S=2}|\bar K^0}\rangle .  \label{eq:m_K-SD}
\end{equation}
The right-hand side of the above equation can be related to the 
matrix element of the $\Delta I=3/2$ operator $O_{\Delta I=3/2}$ in 
the effective weak Hamiltonian, 
\begin{equation}
\langle{K^0|O_{\Delta S=2}|\bar K^0}\rangle 
= \sqrt{2}\langle{\pi^0|O_{\Delta I=3/2}|\bar K^0}\rangle , 
                                          \label{eq:m_K-SU_f(3)}
\end{equation}
in the $SU_f(3)$ symmetry limit\cite{SU_f(3)} or by using the 
asymptotic $SU_f(3)$ symmetry\cite{Delta m_K-asymp} which implies 
a flavor $SU_f(3)$ symmetry of matrix elements of operators (like 
charges, currents, etc.) taken between single hadron states with 
${\bf 1}$-${\bf 8}$ mixing in the IMF\cite{asymptotic-symm}. 

As will be seen later, the short distance amplitudes for the 
$K \rightarrow \pi\pi$ decays do not satisfy the well-known 
$\Delta I=1/2$ rule and their $\Delta I=1/2$ part is much smaller 
than the observed one. However the long distance amplitudes are given 
approximately by asymptotic ground-state-meson matrix elements of
$H_w$ in the present perspective and are expected to be much larger 
than the short distance ones to reproduce the observation. Therefore 
the $\Delta I=1/2$ rule in the $K \rightarrow \pi\pi$ decays will be 
understood rather easily if $(\Delta m_K)_{\rm SD}$ is suppressed. 
On the contrary, if $(\Delta m_K)_{\rm SD}$ dominates $\Delta m_K$, 
it will be hard to explain the observed $\Delta I=1/2$ rule in the 
$K \rightarrow \pi\pi$ decays in a simple way. 

The long distance contribution $(\Delta m_K)_{\rm LD}$ can be written 
in an elegant form in the IMF\cite{Itzykson},
\begin{equation}
 (\Delta m_K)_{\rm LD} = 
\int {dm_n^2 \over 2m_K(m_K^2 - m_n^2)}
        \Bigl\{[\langle{n|H_w|K_L}\rangle ]^2 
                  - [\langle{n|H_w|K_S}\rangle ]^2 \Bigr\}.       
                                             \label{eq:Itzykson}
\end{equation}
It will be dominated by pole contributions of pseudo scalar and 
vector mesons and by $(\pi\pi)$ continuum contributions, 
\begin{equation}
(\Delta m_K)_{\rm LD} 
\simeq (\Delta m_K)_{\rm pole} + (\Delta m_K)_{\pi\pi}. 
                                        \label{eq:Delta m_K-long}
\end{equation}
As the $\pi\pi$ continuum contribution, we here take the following 
value\cite{Pennington}, 
\begin{equation}
{(\Delta m_K)_{\pi\pi} \over \Gamma_{K_S}}
= 0.22 \pm 0.03,                                \label{eq:m_K-pipi}
\end{equation}
which has been obtained by using Omnes-Mushkevili equation and the 
measured $\pi\pi$ phase shifts, where $\Gamma_{K_S}$ denotes the full 
width of $K_S$. Therefore we hereafter can concentrate on the pole 
contribution which is approximated by 
\begin{equation}
(\Delta m_K)_{\rm pole}
\simeq \Biggl\{\sum_{P_i}{|\langle{K_L|H_w|P_i}\rangle|^2 
\over 2m_K(M_K^2 - m_{P_i}^2)}
- \sum_{V_i}{|\langle{K_S|H_w|V_i}\rangle|^2 
\over 2m_K(m_K^2 - m_{V_i}^2)}\Biggr\}
\label{eq:m_K-pole},
\end{equation}
in the IMF, where $P_i=\pi^0$, $\eta$ and $\eta'$ and $V_i=\rho^0$,
$\omega$ and $\phi$. (The $\iota$ contribution has been neglected 
since it is expected to be not very important because of its high
mass.) As seen in Eq.(\ref{eq:m_K-pole}), the pole contribution to 
the $K_L$-$K_S$ mass difference $(\Delta m_K)_{\rm pole}$ has been 
described in terms of asymptotic ground-state-meson matrix elements 
of $H_w$. 

\section{Two pion decays of $K$ mesons}

Amplitudes for two pion decays of $K$ mesons are again classified 
into {\it short distance} and {\it long distance} ones. The former 
will be estimated by using the naive factorization below 
\newpage
\begin{center}
\begin{quote}
{Table~I. Naively factorized amplitudes for the 
$K \rightarrow \pi\pi$ decays where terms proportional to $f_-$ are 
neglected.}
\end{quote}
\vspace{0.5cm}

\begin{tabular}
{l|l}
\hline\hline
$\quad\,\,${\rm Decay}
&\hskip 3.5cm {$\quad M_{\rm SD}\,$}
\\
\hline 
$K_S\, \rightarrow \pi^+\pi^-$
& $-iV_{ud}V_{us}
(G_F/\sqrt{2})a_1f_\pi(m_K^2 - m_\pi^2)f_+^{\pi K}(m_\pi^2)$
\\
$K_S\, \rightarrow \pi^0\,\pi^0$
&  \hspace{4cm} $ 0 \hspace{3cm}\,\,\,\,$
\\
$K^+ \rightarrow \pi^+\pi^0$
& $\,\,\,\, iV_{ud}V_{us}(G_F/2)
(a_1+a_2)f_\pi(m_K^2 - m_\pi^2)f_+^{\pi K}(m_\pi^2)$ 
\\
\hline\hline
\end{tabular}

\end{center}
\vspace{1cm}
while 
the latter is assumed to be dominated by dynamical contributions of 
various hadron states as in the previous sections and is estimated 
later by using a hard pion approximation in the IMF. 

The short distance amplitudes for the $K \rightarrow \pi\pi$ decays 
are estimated by using the naive factorization in the BSW 
scheme\cite{BSW}. As an example, we consider the amplitude for 
the $K^+ \rightarrow \pi^+\pi^0$ decay which is given by 
\begin{eqnarray}
&& M_{\rm SD}(K^+(p) \rightarrow \pi^0(p')\pi^+(q))   \nonumber\\
&&\qquad = {G_F \over \sqrt{2}}V_{us}V_{ud}
\Bigl\{ 
a_1\langle \pi^+(q)|(\bar ud)_{V-A}|0\rangle 
\langle \pi^0(p')|(\bar su)_{V-A}|K^+(p) \rangle         \nonumber\\
&&\hspace{3cm} + a_2\langle \pi^0(p')|(\bar uu)_{V-A}|0\rangle
\langle \pi^+(q)|(\bar sd)_{V-A}|K^+(p) \rangle
\Bigr\}.
                                                    \label{eq:FACT}
\end{eqnarray}
Factorizable amplitudes for the other $K \rightarrow \pi\pi$ decays 
also can be calculated in the same way. To evaluate these amplitudes, 
we use the following parameterization of matrix elements of currents, 
\begin{equation}
\langle \pi(q)|A_\mu^{(\pi)}|0 \rangle 
                         = -if_{\pi}q_\mu, \quad {\rm etc.}, 
                                                      \label{eq:PCAC}
\end{equation}
\begin{eqnarray}
&&\langle \pi(p')|V_\mu^{(\pi K)}|K(p)\rangle 
= (p+p')_\mu f_+^{(\pi K)}(q^2) 
              + q_\mu f_-^{(\pi K)}(q^2), \quad {\rm etc.}, 
                                                        \label{eq:KD}
\end{eqnarray}
where $q = p - p'$. Using these expressions of current matrix
elements, we obtain the factorized amplitudes listed in Table~I, 
where terms proportional to $f_-(q^2)$ have been neglected since 
their coefficients are small in the spectator decays and, in possible 
annihilation decays, they 
are proportional to $a_2$. The penguin 
contribution has also been neglected since, recently, it is 
considered to be very small\cite{BBL} in contrast with the old 
expectation\cite{SVZ}. If the values of $a_1$ and $a_2$ with the 
leading order QCD corrections\cite{BBL} are taken, it will be seen, 
since $|a_1| \gg |a_2|$, that the factorized amplitude for the 
$K^0\rightarrow \pi^0\pi^0$ decay which is described by the color 
mismatched diagram, $\bar s \rightarrow \bar d\,+\,(u\bar u)_1$, is 
proportional to $a_2$ and therefore is much smaller (the color 
suppression) than those for the spectator decays and that the short
distance amplitude for the $K^+\rightarrow\pi^+\pi^0$ decay is
considerably larger than the observed one. For the same reason, 
the size of the $\Delta I=1/2$ amplitude is not much larger than 
the $\Delta I=3/2$ part. Therefore it is hard to reproduce the
well-known approximate $\Delta I=1/2$ rule by the short distance 
amplitudes for the $K \rightarrow \pi\pi$ decays. 

Next we study long distance amplitudes for these decays by using a
hard pion approximation in the IMF, {\it i.e.}, we evaluate the
amplitudes at a slightly unphysical point ${\bf q}\rightarrow 0$ 
in the IMF\cite{hard pion,suppl}. The hard pion amplitude as the long
distance one is written in the form, 
\begin{equation}
M_{\rm LD}(K\rightarrow \pi_1\pi_2) 
\simeq M_{\rm ETC}(K\rightarrow \pi_1\pi_2) 
+ M_{\rm S}(K\rightarrow \pi_1\pi_2),            \label{eq:hard pion}
\end{equation}
where $M_{\rm ETC}$ and $M_{\rm S}$ are given by 
\begin{equation}
M_{\rm ETC}(K\rightarrow \pi_1\pi_2)
= {i \over \sqrt{2}f_{\pi}}
     \langle{\pi_2|[V_{\bar \pi_1}, H_w]|K}\rangle 
                      + (\pi_2 \leftrightarrow \pi_1)   \label{eq:ETC}
\end{equation}
and
\begin{eqnarray} 
&&M_{\rm S}(K\rightarrow \pi_1\pi_2)
= {i \over \sqrt{2}f_{\pi}}
\Biggl\{\sum_n\Bigl({m_{\pi}^2 - m_{K}^2 
                                 \over m_n^2 - m_{K}^2}\Bigr)
  \langle{\pi_2|A_{\bar \pi_1}|n}\rangle
                         \langle{n|H_w|K}\rangle  \nonumber\\
&&\hspace{4.3cm} + \sum_\ell\Bigl({m_{\pi}^2 - m_{K}^2 
                              \over m_\ell^2 - m_{\pi}^2}\Bigr)
\langle{\pi_2|H_w|\ell}\rangle
                  \langle{\ell|A_{\bar \pi_1}|K}\rangle\Biggr\} 
+ (\pi_2 \leftrightarrow \pi_1),  
                                                    \label{eq:SURF}
\end{eqnarray}
respectively, where $[V_\pi + A_\pi, H_w]=0$ has been used. (See 
Refs.\cite{hard pion} and \cite{suppl} for notations.) $M_{\rm ETC}$ 
has the same form as the one in the old soft pion approximation but 
now has to be evaluated in the IMF. The surface term has been given 
by a sum of all possible pole amplitudes, {\it i.e.}, $n$ and $\ell$ 
run over all possible single meson states, not only ordinary 
$\{q\bar q\}$, but also hybrid $\{q\bar qg\}$, four-quark 
$\{qq\bar q\bar q\}$, glue-balls, etc. However, since values of wave 
functions of orbitally excited $\{q\bar q\}_{L \neq 0}$ states at the 
origin are expected to vanish in the non-relativistic quark model, 
and more generally, wave function overlappings between the
ground-state $\{q\bar q\}_0$ and excited-state-meson states are 
expected to be small, we neglect contributions of excited-state 
mesons to the amplitudes except for the $K^+\rightarrow \pi^+\pi^0$ 
in which the ground-state-meson contributions can be strongly 
suppressed because of the (approximate) $\Delta I=1/2$ rule in the 
asymptotic ground-state-meson matrix elements of $H_w$ as will be 
seen later. Asymptotic matrix elements of isospin $V_\pi$ and its 
axial counterpart $A_\pi$ involved in the amplitudes can be well 
parameterized by using (asymptotic) $SU_f(3)$ symmetry. Therefore 
the hard pion amplitude in Eq.(\ref{eq:hard pion}) with 
Eqs.(\ref{eq:ETC}) and (\ref{eq:SURF}) as the non-factorizable long 
distance contribution is approximately described in terms of 
asymptotic ground-state-meson matrix elements of $H_w$. 

Amplitudes for dynamical hadronic processes, in general, can be 
described in the form, 
({\it continuum contribution}) + ({\it  Born term}).  
In the present case, $M_{\rm S}$ is given by a sum of pole amplitudes 
so that $M_{\rm ETC}$ corresponds to the continuum 
contribution\cite{MATHUR} which can develop a phase relative to the 
Born term. Therefore, using isospin eigen amplitudes 
$M_{ETC}^{(I)}$'s and their phases $\delta_I$'s, we here parameterize 
the ETC terms as 
\begin{eqnarray}
&&M_{\rm ETC}(K_S^0\, \rightarrow \pi^+\pi^-) 
= \quad\,\,\,\,{2 \over 3}\,\,M_{\rm ETC}^{(2)}(K\rightarrow \pi\pi)
                                        e^{i\delta_{2}}\, 
+\,\, {1 \over 3}\,\,M_{\rm ETC}^{(0)}(K\rightarrow \pi\pi)
                                        e^{i\delta_{0}},  
                                                \label{eq:ETCMP}\\
&&M_{\rm ETC}(K_S^0\, \rightarrow \pi^0\,\pi^0\,) 
= -{2\sqrt{2} \over 3}M_{\rm ETC}^{(2)}(K\rightarrow \pi\pi)
                                        e^{i\delta_{2}}\, 
+  {\sqrt{1\over 2}}M_{\rm ETC}^{(0)}(K\rightarrow \pi\pi)
                                        e^{i\delta_{0}},  
                                                \label{eq:ETCZZ}\\
&&M_{\rm ETC}(K^+ \rightarrow  \pi^+\pi^0\,) 
= \qquad\quad M_{\rm ETC}^{(2)}(K\rightarrow \pi\pi)
                                     e^{i\delta_{2}}, 
                                                \label{eq:ETCZP}
\end{eqnarray}
since the $S$-wave $\pi\pi$ final states can have isospin $I=0$ and 2. 
Therefore the so-called final state interactions are now included in 
the long distance amplitudes. It is much more natural than the usual
case in which phase factors are multiplied to the factorized
amplitudes by hand. 

In this way, we see that the long distance amplitudes for the 
$K \rightarrow \pi\pi$ decays will satisfy well the $\Delta I = 1/2$ 
rule if the asymptotic ground-state-meson matrix elements of $H_w$ 
satisfy the same rule. Inversely, to reproduce the observed
approximate $\Delta I=1/2$ rule in the $K\rightarrow \pi\pi$
amplitudes, the long distance contributions dominate these decays and 
the ground-state-meson matrix elements of $H_w$ should satisfy the 
same rule. Therefore we here assume that the hard pion amplitude as 
the long distance contribution dominates the $\Delta I=1/2$ amplitude 
in the $K\rightarrow \pi\pi$ decays since the short distance 
$\Delta I=1/2$ amplitude is much smaller than the observed one as 
discussed before and that the asymptotic ground-state-meson matrix 
elements of $H_w$ satisfy the $\Delta I=1/2$ rule. (We will 
demonstrate in the next section, using a simple quark counting, that 
they are obliged to satisfy the $\Delta I=1/2$ rule.) 

By neglecting small contributions of excited states and seemingly 
small $\Delta I=3/2$ asymptotic ground-state-meson 
matrix elements of $H_w$, the long distance amplitudes for the 
$K\rightarrow \pi\pi$ decays can be summarized as follows,
\begin{eqnarray}
&&M_{\rm LD}(K_S^0\, \rightarrow \pi^+\pi^-)
\simeq -{i \over f_\pi}\langle{\pi^+|H_w|K^+}\rangle
\Biggl\{
e^{i\delta_0} + \sqrt{1\over 2}h
{\langle{\pi^+|H_w|K^{*+}}\rangle 
                             \over \langle{\pi^+|H_w|K^+}\rangle}
\Biggr\},                                       \label{eq:LD-pm}\\
&&M_{\rm LD}(K_S^0\, \rightarrow \pi^0\,\pi^0\,)
\simeq -\sqrt{1\over 2}M_{\rm LD}(K_S \rightarrow \pi^+\pi^-), 
                                                \label{eq:LD-00}\\
&&M_{\rm LD}(K^+ \rightarrow \pi^+\pi^0\,) \simeq 0, 
                                                 \label{eq:LD-pz}
\end{eqnarray}
where the size of $h = \langle{\pi^-|A_{\pi^-}|\rho^0}\rangle$ is 
estimated to be $|h|\simeq 1.0$ from the observed decay 
rate\cite{PDG}, 
$\Gamma(\rho \rightarrow \pi\pi)_{\rm expt}\simeq 150$ MeV, 
by using PCAC. In this way, the $K\rightarrow \pi\pi$ amplitudes have
been described approximately by the asymptotic ground-state-meson 
matrix elements of $H_w$ and the iso-scalar $S$-wave $\pi\pi$ phase 
shift $\delta_0$. In Eq.(\ref{eq:LD-pz}), the right hand side is 
vanishing since the excited-state meson contributions and the 
$\Delta I=3/2$ asymptotic ground-state-meson matrix 
elements of $H_w$ have been neglected. As will be seen later, the 
$\Delta I = 3/2$ part of the long distance amplitude can be supplied 
through four-quark meson pole amplitudes even if the asymptotic 
ground-state-meson matrix elements of $H_w$ satisfy the 
$\Delta I=1/2$ rule. They can interfere destructively with the too 
big $\Delta I = 3/2$ part of the factorized amplitudes in Table~I. 

\section{Parameterization of asymptotic matrix elements of $H_w$}

We have described approximately the long distance amplitudes for 
the $K\rightarrow \pi\pi$ decays, the $K^0$-$\bar K^0$ mixing and 
the $K_L\rightarrow \gamma\gamma^{(*)}$ using the asymptotic 
ground-state-meson matrix elements of $H_w$ and have seen that the 
$\Delta I=1/2$ rule in the $K\rightarrow \pi\pi$ decays is mainly 
controlled by the same selection rule in the asymptotic 
ground-state-meson matrix elements of $H_w$. We have also seen that 
$(\Delta m_K)_{\rm SD}$ is related to 
$\langle{\pi^0|O_{\Delta I=3/2}|\bar K^0}\rangle$ 
and hence the former should vanish if the 
ground-state-meson matrix elements of $H_w$ satisfy the 
$\Delta I=1/2$ rule. 

Before we parameterize asymptotic matrix elements of $H_w$, we study 
constraints on them using a simple quark counting\cite{TBD}. 
The effective weak Hamiltonian $H_w$ has been given by a sum of four 
quark operators $O_{\pm}$ (and the penguin operator which always 
satisfies the $\Delta I=1/2$ rule) in Eq.(\ref{eq:HW-pm}). The normal 
ordered four-quark operators $O_\pm$ can be expanded into a sum of 
products of (a) two creation operators to the left and two
annihilation operators to the right, (b) three creation operators to 
the left and one annihilation operator to the right, (c) one creation 
operator to the left and three annihilation operators to the right, 
and (d) all (four) creation operators or annihilation operators of 
quarks and anti-quarks. We associate (a)$-$(d) with quark-line 
diagrams describing different types of matrix elements of $O_\pm$. 
For (a), we utilize the two creation and annihilation operators to 
create and annihilate, respectively, the quarks and anti-quarks 
belonging to the meson states $|\{q\bar q\}\rangle$ and 
$\langle \{q\bar q\}|$ in the asymptotic matrix elements of $O_\pm$. 
For (b) and (c), we need to add a spectator quark or anti-quark to 
reach {\it physical} processes, 
$\langle {\{qq\bar q\bar q\}|O_\pm|\{q\bar q\}}\rangle$ and 
$\langle {\{q\bar q\}|O_\pm|\{qq\bar q\bar q\}}\rangle$,  
where $\{qq\bar q\bar q\}$ denotes four-quark mesons\cite{Jaffe}. 
They can be classified into the following four types, 
$\{qq\bar q\bar q\} = [qq][\bar q\bar q] \oplus (qq)(\bar q\bar q) 
\oplus \{[qq](\bar q\bar q) \pm (qq)[\bar q\bar q]\}$,  
where () and [] denote symmetry and antisymmetry, respectively, under
the exchange of flavors between them. We here consider contributions 
only of the first two since the others do not have 
$J^{P(C)}=0^{+(+)}$. 

While we count all possible connected quark-line diagrams, we forget 
color degree of freedom of quarks since they will be compensated by a 
sea of soft gluons carried by light mesons and have to be careful 
with the order of the quark(s) and anti-quarks(s) in $O_\pm$ since 
symmetry (or antisymmetry) property of wave functions of meson states 
sandwiching $O_\pm$ under exchanges of quark and anti-quark plays an 
important role. Noting that the wave function of the ground-state 
$\{q\bar q\}_0$ meson is antisymmetric under the exchange of its 
quark and anti-quark\cite{CLOSE}, we obtain the following constraints 
on asymptotic matrix elements of $O_\pm$\cite{TBD},
\begin{eqnarray}
&&\langle{\{q\bar q\}_0|O_+|\{q\bar q\}_0}\rangle = 0, 
                                                \label{eq:SUM-G}\\
&&\langle{[qq][\bar q\bar q]|O_+|\{q\bar q\}_0}\rangle
= \langle{\{q\bar q\}_0|O_+|[qq][\bar q\bar q]}\rangle = 0, 
                                             \label{eq:SUM-anti}\\
&&\langle{(qq)(\bar q\bar q)|O_-|\{q\bar q\}_0}\rangle 
= \langle{\{q\bar q\}_0|O_+|(qq)(\bar q\bar q)}\rangle =0. 
                                                 \label{eq:SUM-sym}
\end{eqnarray}
The above equations imply that the asymptotic ground-state-meson
matrix elements of $H_w$ satisfy the $\Delta I=1/2$ rule and its
violation in the long distance amplitudes for the 
$K\rightarrow \pi\pi$ decays can be supplied through the four-quark 
$(qq)(\bar q\bar q)$ meson contributions which can interfere 
destructively with the too big $\Delta I = 3/2$ part of the 
factorized amplitude in Table~I. However, since our purpose in this 
paper is not to discuss the $\Delta I = 1/2$ rule and its violation, 
we do not consider them any more. The same quark counting leads 
directly to 
$\langle{K^0|O_{\Delta S=2}|\bar K^0}\rangle 
= 0$\cite{Terasaki-FF,Delta m_K-asymp} 
which is compatible with Eq.(\ref{eq:SUM-G}) as discussed before. 

Now we are ready to parameterize the asymptotic ground-state-meson 
matrix elements of $H_w$. To reproduce the observed $\Delta I=1/2$ 
rule in the $K\rightarrow \pi\pi$ decays, we need the 
$\Delta I=1/2$ rule for the ground-state-meson matrix elements of 
$H_{w}$ with a sufficient precision. It is all right if one accepts
the above quark counting. (If not, one has to assume the 
$\Delta I=1/2$ rule for the ground-state-meson matrix elements of 
$H_{w}$.) Anyway, neglecting seemingly small (or {\it zero} in the 
above quark counting) $\Delta I=3/2$ contributions, we parameterize 
the ground-state-meson matrix elements of $H_{w}$ as follows,  
\hfil\break 
(A) helicity $\lambda=0$ matrix elements:
\begin{eqnarray}
&&\langle{\pi^-|H_{w}|K^-}\rangle 
= \langle{\pi^-|H_{w}|K^{*-}}\rangle 
= \langle{\rho^-|H_{w}|K^-}\rangle = \qquad\,\, (1 + r_{0})H_{0},   
                                                   \label{eq:pi^+}\\
&&\langle{\pi^0\,|H_{w}\,|\bar K^0\,}\rangle 
= \langle{\pi^0\,|H_{w}|\bar K^{*0}}\rangle \,
= \langle{\rho^0\,|H_{w}|\bar K^0\,}\rangle 
= -\sqrt{1\over 2}(1 + r_{0})H_{0},   
                                                   \label{eq:pi^0}\\
&&\langle{\eta_{0}\,\,|H_{w}\,|\bar K^0\,}\rangle 
= \langle{\eta_{0}\,\,|H_{w}|\bar K^{*0}}\rangle 
= \langle{\,\omega\,\,|H_{w}|\bar K^0\,}\rangle 
= -\sqrt{1\over 2}(1 - r_{0})H_{0},   
                                                   \label{eq:eta_{0}}\\
&&\langle{\eta_{s}\,\,|H_{w}\,|\bar K^0\,}\rangle 
= \langle{\eta_{s}\,\,|H_{w}|\bar K^{*0}}\rangle 
=\, \langle{\,\phi\,\,|H_{w}|\bar K^0\,}\rangle 
=\qquad\qquad\,\,\,\,  r_{0}H_{0},   
                                                   \label{eq:eta_{s}}
\end{eqnarray}
(B) helicity $\lambda=\pm1$ matrix elements:
\begin{eqnarray}
&& \langle{\rho^0|H_{w}|\bar K^{*0}}\rangle_{\pm 1}
= -\sqrt{1\over 2}(1 + r_{1})H_{1},   
                                               \label{eq:rho^0}\\
&& \langle{\,\omega\,|H_{w}|\bar K^{*0}}\rangle_{\pm 1} 
= -\sqrt{1\over 2}(1 - r_{1})H_{1},            \label{eq:omega}\\
&& \langle{\,\phi\,\,|H_{w}|\bar K^{*0}}\rangle_{\pm 1} 
= \qquad\qquad\,\,\,\, r_{1}H_{1},            \label{eq:phi}
\end{eqnarray}
where iso-singlet pseudo scalar mesons $\eta$ and $\eta'$ are
written as 
\begin{eqnarray}
&&\eta = \Biggl(\sqrt{1\over 3}{\rm cos}{\theta_P} 
            \, - \sqrt{2\over 3}{\rm sin}{\theta_P}\Biggr)\eta_0 
     - \Biggl(\quad\sqrt{2\over 3}{\rm cos}{\theta_P} 
             + \sqrt{1\over 3}{\rm sin}{\theta_P}\Biggr)\eta_s, 
                                                \label{eq:eta}\\
&&\eta'= \Biggl(\sqrt{1\over 3}{\rm sin}{\theta_P} 
              + \sqrt{2\over 3}{\rm cos}{\theta_P}\Biggr)\eta_0 
    + \Biggl(-\,\sqrt{2\over 3}{\rm sin}{\theta_P} 
              + \sqrt{1\over 3}{\rm cos}{\theta_P}\Biggr)\eta_s, 
                                                 \label{eq:eta'} 
\end{eqnarray}
in terms of their components 
$\eta_{0} \sim (u\bar u + d\bar d)/\sqrt{2}$ and 
$\eta_s \sim (s\bar s)$. 
The mixing angle is usually taken to be 
$\theta_P \simeq -20^\circ$\cite{PDG}. The $\omega$-$\phi$ mixing has
been assumed to be ideal. The parameters $r_0$ and $r_1$ denote 
contributions of the penguin relative to $O_-$ in the helicity 
$\lambda=0$ and $\lambda=\pm 1$ matrix elements of $H_w$, 
respectively. $H_0$ and $H_1$ provide their normalizations. We 
have parameterized the asymptotic matrix elements of $H_w$ between 
pseudo scalar and vector meson states in the same manner. It can be 
justified by a simple algebraic procedure\cite{hard pion,suppl} 
(spins are not very important in the IMF). In (B), the helicity 
$\lambda=\pm 1$ matrix elements, $r_1$ will be neglected hereafter 
since it is expected to be small because of the small coefficient of 
the penguin and because of a helicity consideration. 

\section{Comparison with experiments}

Inserting the above parameterization into the amplitudes given in 
the previous sections, we can compare our result with experiments. 
Since our result contains many parameters, however, we here 
estimate them by using various experimental data. 

Sizes of the amplitudes $A(P_i \rightarrow \gamma\gamma)$'s in 
Eq.(\ref{eq:P-pole}) can be estimated from the measured values of the 
decay rates\cite{PDG}, 
$\Gamma(\pi^0 \rightarrow \gamma\gamma)_{\rm expt} 
= (7.7 \pm 0.6)\,\,{\rm eV}$, 
$\Gamma(\eta \rightarrow \gamma\gamma)_{\rm expt} 
= (0.46 \pm 0.04)\,\,{\rm keV}$ and  
$\Gamma(\eta' \rightarrow \gamma\gamma)_{\rm expt} 
= (4.26 \pm 0.19)\,\,{\rm keV}$.  
Their relative signs are taken to be compatible with the quark
model. The $V$-$V'$-$P$, ($V,\,V'=K^*,\,\rho,\,\omega$ and $\phi$;
$P= K,\,\pi,\,\eta$ and $\eta'$), coupling constants can be estimated 
from the observed rates for the radiative decays of $K^*$ by using 
$SU_f(3)$ symmetry and the VMD with the $\gamma$-$V$ coupling 
strengths\cite{Terasaki-VMD}, 
$X_{\rho}(0)= 0.033 \pm 0.003$ (GeV)$^2$, 
$X_{\omega}(0)= 0.011 \pm 0.001$ (GeV)$^2$ and 
$X_{\phi}(0)= -0.018 \pm 0.001$ (GeV)$^2$, 
estimated from experiments on photo-productions of vector mesons. 
Although these coupling strength can have momentum square ($k^2$) 
dependence, we neglect it in this paper since they are mild in the
region $k^2 < m_K^2$. From 
$\Gamma(K^{*0} \rightarrow K^0\gamma)_{\rm expt} 
= (0.115 \pm 0.012)$ MeV\cite{PDG}, 
we obtain $|G_{K^{*0}K^0\rho^0}| \simeq 0.856$ (GeV)$^{-1}$ and then 
$G_{\omega\pi^0\rho^0} = -2G_{K^{*0}K^0\rho^0}$ using$SU_f(3)$. 
In this way, we can reproduce well the observed rate, 
$\Gamma(\pi^0\rightarrow \gamma\gamma)_{\rm expt}$. The value of the 
matrix elements of axial charges can be estimated to be 
$|h| = |\langle{\pi^-|A_{\pi^-}|\rho^0}\rangle|\,\,
\bigl(\,=\sqrt{2}|\langle{K^+|A_{\pi^+}|K^{*0}}\rangle|\bigr) 
\simeq 1.0$ 
from the observed rate\cite{PDG}, 
$\Gamma(\rho \rightarrow \pi\pi)_{\rm expt} \simeq 150$ MeV, 
by using PCAC and (asymptotic) $SU_f(3)$. The above value of $|h|$ 
can reproduce considerably well 
$\Gamma(K^* \rightarrow K\pi)_{\rm expt}$. 
The size of $\langle{\pi|H_w|K}\rangle$ 
can be estimated from the observed rate for the 
$K_S \rightarrow \pi^+\pi^-$ decay by using Eq.(\ref{eq:LD-pm}) with 
$\langle{\pi|H_w|K}\rangle = \langle{\pi|H_w|K^*}\rangle$ 
and the $S$-wave $\pi\pi$ phase shift $\delta_0 \simeq (50-60)^\circ$ 
at $m_K$\cite{Kamal} and by taking into account the small 
contribution of the factorized amplitude for the same decay in 
Table~I; 
\begin{equation}
|\langle{\pi^+|H_w|K^+}\rangle|\,
=\,|\langle{\pi^+|H_w|K^{*+}}\rangle|\,
\simeq 1.69\times 10^{-7}m_K^2.            \label{eq:matrix element}
\end{equation}

The parameters which are included in the amplitudes given in 
the previous sections but still have not been estimated are 
$\alpha_{K^*} 
= \langle{\rho^+|H_w|K^{*+}}\rangle_{\pm 1}
/\langle{\pi^+|H_w|K^+}\rangle$ 
and $r_0$ (describing the contribution of the penguin relative to 
$O_-$ in the asymptotic ground-state-meson matrix elements of $H_w$ 
with the helicity $\lambda=0$). We now search for values of these 
parameters to reproduce $(\Delta m_K)_{\rm expt}$ and 
$\Gamma(K_L \rightarrow \gamma\gamma)_{\rm expt}$ mentioned before. 
[We have already used 
$\Gamma(K_S^0 \rightarrow \pi^+\pi^-)_{\rm expt}$ 
to estimate the size of $\langle{\pi^+|H_w|K^+}\rangle$.] 
For the $\Delta m_K$, relative importance between 
$(\Delta m_K)_{\rm SD}$ and $(\Delta m_K)_{\rm LD}$ is still not 
known. However, if we accept the result from the quark counting 
presented in the previous section, we have $(\Delta m_K)_{\rm SD}=0$ 
and we can rather easily understand the observed $\Delta I=1/2$ rule 
in the $K\rightarrow \pi\pi$ decays. In this case (i), the pole 
contribution which we have calculated in {\bf 3} should be compared 
with 
\begin{eqnarray}
&&{(\Delta m_K)_{\rm pole}\over\Gamma_{K_S}}\,
\simeq \quad\Bigl({\Delta m_K\over\Gamma_{K_S}}\Bigr)_{\rm expt}\quad 
-\quad {(\Delta m_K)_{\pi\pi}\over\Gamma_{K_S}} \nonumber\\
&&\hspace{2cm}\,\,\simeq (0.477 \pm 0.002)\,\,\,
-\,\,\,(0.22 \pm 0.03), 
                                             \label{eq:m_K-pole-num}
\end{eqnarray}
where the value of $(\Delta m_K)_{\pi\pi}/\Gamma_{K_S}$ has been given 
in Ref.\cite{Pennington} as mentioned before. Inserting the
parameterization of the asymptotic ground-state-meson matrix elements
of $H_w$ in the previous section into $(\Delta m_K)_{\rm pole}$ in 
Eq.(\ref{eq:m_K-pole}) and $A(K_L \rightarrow \gamma\gamma)$ in 
Eq.(\ref{eq:2-gamma}), we find two possible solutions, 
\begin{eqnarray}
&&{\rm (a)}\,\,0.31 < r_0 < 0.35 \quad {\rm and} \quad
        1.05 < \alpha_{K^*} < 1.20,    \label{eq:sol-a}\\
&&{\rm (b)}\,\,0.31 < r_0 < 0.35 \quad {\rm and} \quad
        3.40 < \alpha_{K^*} < 3.55,    \label{eq:sol-b}
\end{eqnarray}
which can reproduce the value of 
$(\Delta m_K)_{\rm pole}/\Gamma_{K_S}$ in Eq.(\ref{eq:m_K-pole-num}) 
and $\Gamma(K_L\rightarrow \gamma\gamma)_{\rm expt}$, where 
$|A_P(K_L\rightarrow \gamma\gamma)| 
> |A_{K^*}(K_L\rightarrow \gamma\gamma)|$ for (a) and 
$|A_P(K_L\rightarrow \gamma\gamma)| 
< |A_{K^*}(K_L\rightarrow \gamma\gamma)|$ for (b), 
respectively. 

Next, we consider the case (ii) in which  $(\Delta m_K)_{\rm SD}$ 
dominates the $K_L$-$K_S$ mass difference, {\it i.e.}, 
$(\Delta m_K)_{\rm LD}= 0$. In this case, it is not very easy to 
understand the $\Delta I=1/2$ rule in the $K\rightarrow \pi\pi$ 
decays since 
$\langle{\pi|O_{\Delta I=3/2}|K}\rangle \neq 0$. 
Using Eq.(\ref{eq:m_K-SU_f(3)}) and the observed value of 
$\Delta m_K$, we obtain, 
$\langle{\pi^0|H_w^{(\Delta I=3/2)}|K^0}\rangle 
\simeq 0.22 \times 10^{-7} m_K^2$, 
which is considerably smaller than the size of 
$\langle{\pi^+|H_w|K^+}\rangle$ estimated phenomenologically in 
Eq.(\ref{eq:matrix element}) by neglecting the $\Delta I=3/2$ part 
of asymptotic ground-state-meson matrix elements of $H_w$, 
{\it i.e.}, 
\begin{equation}
\Biggl|\sqrt{1 \over 2}
{\langle{\pi^+|H_w^{(\Delta I=1/2)}|K^+}\rangle 
    \over \langle{\pi^0|H_w^{(\Delta I=3/2)}|K^0}\rangle}\Biggr| 
\simeq 5.4 
\end{equation}
while the ratio of the coefficient $c_-$ to $c_+$ is 
${c_-/ c_+} \simeq 4.3$. Then the long distance amplitude 
$M_{\rm LD}(K^+\rightarrow \pi^+\pi^0)$ which will be proportional 
to $\langle{\pi|H_w^{(\Delta I=3/2)}|K}\rangle$ if contributions of 
excited-meson-states are neglected can interfere destructively with 
too big $M_{\rm SD}(K^+\rightarrow \pi^+\pi^0)$. A 
\newpage
\begin{center}
\begin{quote}
{Table~II. Branching fractions for the Dalitz decays of $K_L$. The
values of the parameters in (a), (b) and (c) are taken from the
corresponding solutions in the text. Data values with ($\ast$), 
($\dag$) and ($\ddag$) are taken from Refs. \cite{PDG}, \cite{E799} 
and \cite{NA48}, respectively. 
}
\end{quote}
\vspace{0.5cm}

\begin{tabular}
{c|c|c|c}
\hline\hline
\multicolumn{2}{c|}
{}
&$R_{\gamma e^+e^-}(\times 10^{-2})$ 
&$R_{\gamma\mu^+\mu^-}(\times 10^{-4})$
\\
\hline 
(i)\,\,$(\Delta m_K)_{\rm SD}=0$
&\begin{tabular}{c}
${\rm (a)}\,\, r_0 = 0.330,\,\alpha_{K^*} = 1.13$\\
${\rm (b)}\,\,r_0 = 0.330,\,\alpha_{K^*} =  3.47$
\end{tabular}
& \begin{tabular}{c}
$1.6$\\
$1.7$
\end{tabular}
& \begin{tabular}{c}
$5.6$\\
$6.6$
\end{tabular}
\\
\hline
(ii)\,$(\Delta m_K)_{\rm LD}=0$
&${\rm (c)}\,\,r_0 = 0.139,\,\alpha_{K^*} =  4.32$
& $1.7$
& $6.8$
\\
\hline
\multicolumn{2}{c|}
{Experiments}
& $(1.6 \pm 0.1)(\ast)$
& \begin{tabular}{c}
$(5.6 \pm 0.8)(\dag)$\\
$(5.9 \pm 1.8)(\ddag)$
\end{tabular}
\\
\hline\hline
\end{tabular}

\end{center}
\vspace{1cm}
sum of these two 
amplitudes leads to 
$\Gamma(K^+\rightarrow \pi^+\pi^0) \simeq 0.14 \times 10^8$ s$^{-1}$ 
which should be compared with 
$\Gamma(K^+\rightarrow \pi^+\pi^0)_{\rm expt} 
\simeq 0.17 \times 10^8$ s$^{-1}$. 
Anyway, we neglect the {\it small} $\Delta I =3/2$ part in the
asymptotic ground-state-meson matrix elements of $H_w$ and then we 
can use the parameterization of them in the previous section and the 
value of $|\langle{\pi^+|H_w|K^+}\rangle|$ in 
Eq.(\ref{eq:matrix element}) as its approximate value. Since 
$(\Delta m_K)_{\rm pole}/\Gamma_{K_S}$ should be cancelled by 
$(\Delta m_K)_{\pi\pi}/\Gamma_{K_S}$ in this case (ii), we put 
\begin{equation}
{(\Delta m_K)_{\rm pole}\over \Gamma_{K_S}} 
= -{(\Delta m_K)_{\pi\pi}\over \Gamma_{K_S}}
= -(0.22 \pm 0.03). 
\end{equation}
Then we find another possible solution which can reproduce the value 
of $(\Delta m_K)_{\rm pole}/\Gamma_{K_S}$ in the above equation and 
$\Gamma(K_L\rightarrow \gamma\gamma)_{\rm expt}$, 
\begin{equation}
{\rm (c)}\,\,0.13 < r_0 < 0.15  \quad {\rm and}\quad 
      4.25 < \alpha_{K^*} < 4.39.                \label{eq:sol-c}
\end{equation}

Inserting the above sets (a) -- (c) of values of $r_0$ and 
$\alpha_{K^*}$ into Eq.(\ref{eq:Dalitz-FF}), we obtain three 
different results on the form factor for the Dalitz decays of $K_L$. 
For experimental data on the form factor, there exist three different 
data, {\it i.e.}, two of them are from the $\gamma e^+e^-$ final 
states\cite{NA31,E845} and the other is from the 
$\gamma\mu^+\mu^-$\cite{E799}. The existing data from different types 
of the final states are not consistent with each other near the 
$\gamma\mu^+\mu^-$ threshold. Our results from the solutions (b) and 
(c) are not very far from the data from the $\gamma e^+e^-$ final 
states but not consistent with the data from the $\gamma\mu^+\mu^-$ 
final states while the one from the solution (a) is close to the data 
from the $\gamma\mu^+\mu^-$ final states near the threshold of the 
$K_L\rightarrow\gamma\mu^+\mu^-$. At higher 
$x = k^2/m_K^2 \,(> 0.4)$, all the three results are consistent with 
almost all the data within their large errors. 

Substituting the above results on the form factor for the Dalitz 
decays of $K_L$ into the formula Eq.(\ref{eq:Dalitz-rate}) with 
Eq.(\ref{eq:Diff-rate}), we can calculate their branching 
fractions\cite{Erratum} as listed in Table II. The rate for the 
Dalitz decay of $K_L$ is mainly determined by the values of the 
form 
factor near the threshold. Therefore, the rate 
$\Gamma(K_L \rightarrow \gamma e^+e^-)$ is not very useful to 
discriminate different theories since its threshold is close to $x=0$ 
where the form factor is usually normalized to be $f(0) = 1$. However 
the threshold of the $K_L \rightarrow \gamma \mu^+\mu^-$ decay is 
considerably distant from the normalization point $x = 0$. 
Therefore, we may discriminate the above three different solutions 
using this decay since they give considerably different values of the 
form factor around the $\gamma\mu^+\mu^-$ threshold. In Table~II, 
it is seen that the data from the $\gamma \mu^+\mu^-$ final state 
seem to favor the solution (a). However, at the present stage where
theoretical and experimental ambiguities are still large, it is hard 
to say definitely what solution is the best. 

\section{Summary}

We have investigated $\Delta m_K$, $K \rightarrow \pi\pi$, 
$K_L \rightarrow \gamma\gamma$ and the Dalitz decays of $K_L$ 
systematically. For $\Delta m_K$, we have considered two extreme 
cases, {\it i.e.}, (i) $(\Delta m_K)_{\rm SD}=0$ and 
(ii) $(\Delta m_K)_{\rm LD}=0$, since we do not know relative weight 
between $(\Delta m_K)_{\rm SD}$ and $(\Delta m_K)_{\rm LD}$ in 
$\Delta m_K$. We have searched possible solutions to reproduce the 
{\it data} on $\Delta m_K$ and $\Gamma(K_L \rightarrow \gamma\gamma)$ 
simultaneously and found two possible solutions, (a) and (b), in the 
case (i) and a possible solution, (c), in the case (ii). Then, using 
these solutions, we have calculated the form factor for the Dalitz 
decays of $K_L$ and their decay rates. Our predictions have been 
compared with the existing experimental data. However, the existing 
data on the form factor from the $K_L \rightarrow \gamma e^+e^-$ 
decay\cite{NA31,E845} and the $K_L \rightarrow \gamma \mu^+\mu^-$ 
decay\cite{E799} are not consistent with each other near the 
$\gamma \mu^+\mu^-$ threshold. The results from two of our solutions, 
(b) and (c), are almost consistent with the data from the 
$\gamma e^+e^-$ final state but considerably higher than the data 
from the $\gamma \mu^+\mu^-$ final state around its threshold. 
The form factor predicted by the solution (a) is close to the data
from the $\gamma \mu^+\mu^-$ final states. The rate 
$\Gamma(K_L \rightarrow \gamma \mu^+\mu^-)$ will be useful to 
discriminate these three solutions in contrast with 
$\Gamma(K_L \rightarrow \gamma e^+e^-)$. The data from 
E799\cite{E799} seems to favor the solution (a) and hence a 
suppression of $(\Delta m_K)_{\rm SD}$ which can explain rather easily 
the $\Delta I=1/2$ rule in the $K \rightarrow \pi\pi$ amplitudes
and can be derived by a simple quark counting. However, it is hard to
conclude definitely the above statement since theoretical and 
experimental ambiguities are still too large. Therefore more 
theoretical and experimental investigations of the Dalitz decays of 
$K_L$ will be needed. 
\vspace{0.5cm}

The author thanks Dr.~K.~E.~Ohl, Dr.~H.~Rohrer, Dr.~D.~Coward and 
Dr.~T.~Nakaya for sending their data values of the form factor for the 
Dalitz decays. He also appreciate Dr.~P.~Singer for arguments against
the $K^*$ pole contribution to the $K_L\rightarrow\gamma\gamma$
decay. 



\begin{references}
\bibitem{QCD-corr} Although the next-leading-order corrections have 
already been calculated in Ref.\cite{BBL}, the result seems to be 
sensitive to a choice of renormalization scheme in contrast with the 
leading order corrections. We here take the latter since we do not
need very precise values of $a_1$ and $a_2$.
\bibitem{asymptotic-symm} A comprehensive review on the asymptotic
flavor symmetry and its fruitful results has been provided in
Ref.\cite{suppl}.
\bibitem{Erratum} Our previous results\cite{Terasaki-FF} on the 
branching fraction, $R_{\gamma e^+e^-}$,  were too small since the 
singular behavior of its kinematical factors near the threshold was
not sufficiently taken into account. They have been improved in 
Table~II.
\end{references}
\end{document}